\documentstyle[12pt,aasms4]{article}
\begin{document}

\newcommand{\etal}{ {\it et al.}}
\def\cen{\centerline}

\def\loe{\lower 0.6ex\hbox{${}\stackrel{<}{\sim}{}$}}
\def\goe{\lower 0.6ex\hbox{${}\stackrel{>}{\sim}{}$}}

\def\ggg{$\gamma$}

%
\def\jref#1 #2 #3 #4 {{\par\noindent \hangindent=3em \hangafter=1
      \advance \rightskip by 0em #1, {\it#2}, {\bf#3}, #4.\par}}
\def\rref#1{{\par\noindent \hangindent=3em \hangafter=1
      \advance \rightskip by 0em #1.\par}}

\def\p{\phantom{1}}
\def\pmu{\mox{$^{-1}$}}
\def\ApJ{{\it Ap.\,J.\/}}
\def\ApJL{{\it Ap.\,J.\ (Letters)\/}}
\def\ApJS{{\it Astrophys.\,J.\ Supp.\/}}
\def\AJ{{\it Astron.\,J.\/}}
\def\AAL{{\it Astr.\,Astrophys.\ Letters\/}}
\def\AAS{{\it Astr.\,Astrophys.\ Suppl.\,Ser.\/}}
\def\MN{{\it Mon.\,Not.\,R.\,Astr.\,Soc.\/}}
\def\Na{{\it Nature \/}}
\def\SAIt{{\it Mem.\,Soc.\,Astron.\,It.\/}}
\def\BGD{\begin{description}}
\def\EDD{\end{description}}
\def\BGF{\begin{figure}}
\def\EDF{\end{figure}}
\def\BGC{\begin{center}}
\def\EDC{\end{center}}
\def\BGT{\begin{tabular}}
\def\EDT{\end{tabular}}
\def\BGE{\begin{equation}}
\def\EDE{\end{equation}}
\def\REFFF{\par\noindent\hangindent 20pt}
\def\DS{\displaystyle}
\def\kms{km^s$^{-1}$}
\def\sbu{mag^arcsec${{-2}$}}
\def\e{\mbox{e}}
\def\dex{\mbox{dex}}
\def\L{\mbox{${\cal L}$}}
 

\newcommand{\porb}{ P_{orb} } 
\newcommand{\Po}{$ P_{orb} \su$}
\newcommand{\pdot}{$ \dot{P}_{orb} \,$}

\newcommand{\pot}{$ \dot{P}_{orb} / P_{orb} \su $}
\newcommand{\s}{ \\ [.15in] }

\newcommand{\mm}{$ \dot{m}$ }
\newcommand{\mdot}{$ |\dot{m}|_{rad}$ }

\newcommand{\myr}{ \su M_{\odot} \su \rm yr^{-1}}
\newcommand{\msol}{\, M_{\odot}}
\newcommand{\ppp}{ \dot{P}_{-20} }

\newcommand{\ci}[1]{\cite{#1}}
\newcommand{\bb}[1]{\bibitem{#1}}

\newcommand{\ch}[1]{\vskip .3in \noindent {\bf #1} \para}

\newcommand{\cms}{ \rm \, cm^{-2} \, s^{-1} }

\newcommand{\nn}{\noindent}

%
%

\renewcommand{\thebibliography}[1]{
  \list
  {[\arabic{enumi}]}{\settowidth\labelwidth{[#1]}\leftmargin\labelwidth
    \advance\leftmargin\labelsep
    \usecounter{enumi}}
    \def\newblock{\hskip .11em plus .33em minus .07em}
    \parsep -2pt
    \itemsep \parsep
    \sloppy\clubpenalty4000\widowpenalty4000
    \sfcode`\.=1000\relax}

\newcommand{\pdott}{ \left( \frac{ \dot{P}/\dot{P}_o}{P_{1.6}^{3}} \right)}

\newcommand{\asca}{{\it ASCA} }
\newcommand{\psr}{PSR~B1259-63 }
\newcommand{\psrp}{PSR~B1259-63} 

\newcommand{\syn}{synchrotron }

\def\egr{ GRO~J1838--04 }
\def\egrp{ GRO~J1838--04}

\def\pt{$10}
\def\cms{ \, {\rm ph \, cm^{-2} \, s^{-1}}$ }

\def\egret{EGRET }
\renewcommand{\etal}{{\it et al.}}
 
\baselineskip 20pt

\nn
\nn

\vspace*{0.5in}

\vskip 0.5in
\title{Discovery of a non-blazar gamma-ray transient
 near the Galactic plane: \egr}

\author{M. Tavani$^1$, R. Mukherjee$^{2,3}$, J.R.  Mattox$^4$, J. Halpern$^1$}
\author{D.J. Thompson$^5$, G. Kanbach$^6$, W. Hermsen$^7$, 
S.N. Zhang$^8$, R. S. Foster$^9$}

\vskip 2.in
{\small 

\nn
1. Columbia Astrophysics Laboratory, Columbia University, New York, NY 10027.\\
2. USRA, NASA/GSFC, Code 610.3, Greenbelt, MD 20771.\\
3. Physics Dept., McGill University, 3600 University St., Montreal, Que, H3A 2T8, Canada.\\
4. Astronomy Dept., Boston University, 725 Commonwealth Avenue, Boston, MA 02215.\\ 
5.  Code 661, NASA Goddard Space Flight Center, Greenbelt, MD 20771.\\
6.  MPE, Giessenbachstrasse, Garching bei Munchen, D-85748, Germany.\\
7. SRON-Utrecht, Sorbonnelaan 2, 3584 CA Utrecht, The Netherlands.\\
8. Universities Space Research Association, NASA/MSFC, Huntsville, AL, 35812.\\
9. Code 7210, Naval Research Laboratory, Washington, DC  20375.

}

\newpage

\vskip .4in
\begin{abstract}
We report the  discovery   of a remarkable 
gamma-ray transient source near the Galactic plane,
\egrp.
This source was serendipitously discovered by EGRET in June 1995
with a peak intensity of
$\sim (4 \pm 1) \times 10^{-6} \rm \, ph \, cm^{-2} \, s^{-1}$
(for photon energies larger than 100~MeV) and a 5.9$\sigma$
significance. At that time, \egr was the second brightest \ggg-ray source
in the sky.
A subsequent  EGRET pointing in late September 1995 
detected the source at a flux smaller than its peak value
 by a factor of $\sim 7$.
We determine  that no radio-loud spectrally-flat
 blazar  is within the  
 error box of \egrp.
We discuss the origin of the \ggg-ray transient source and
 show that 
 interpretations in terms of
AGNs or isolated pulsars are highly  problematic.
\egr
provides strong evidence for the existence of a new class of
 variable gamma-ray sources.
\end{abstract}
\vskip .3in

\vskip .3in
\centerline{Submitted to the {\it Astrophysical Journal Letters}: December 16, 1996}
\centerline{Accepted: February 3, 1997}

\newpage

\section{Introduction}

{ EGRET}  has discovered
$\sim $40 unidentified gamma-ray  sources concentrated near the
Galactic plane (Thompson \etal\ 1995, 1996).
The  nature of these unidentified \ggg-ray sources
is currently unknown.
Of particular interest are time-variable  \egret
sources
with no obvious  radio-loud
blazar  within their error boxes.
Isolated pulsars 
similar to the Crab, Vela, and Geminga pulsars
have an approximately  constant \ggg-ray  flux 
(e.g., Ramanamurthy \etal\ 1995)
and cannot account for the existence of variable sources.
 
We report here the discovery of a strongly variable  \ggg-ray 
source, \egrp.
This source was detected
during an observation of a field which included the Galactic plane 
(Galactic longitude between $l=17^{\circ}$ 
and $l=32^{\circ}$).  This field contains other 
  EGRET sources studied by our group, 2EG~J1813--12 and 2EG~J1828+01
(Tavani \etal\ 1997).
Our search for time variable \ggg-ray
sources near the Galactic plane is 
motivated by models of high-energy emission from Galactic sources
(Tavani 1995).
 However, the possible detection by EGRET of
 extragalactic objects  near the Galactic
plane  cannot {\it a priori} be excluded.
We consider here blazars as plausible candidates of \ggg-ray emission
(e.g., Montigny \etal\ 1995, Hartman \etal\ 1996, Mattox \etal\ 1996,
 hereafter M96).

\vskip .2in
\section{{\it Compton} GRO observations of \egr}

The source \egr  was
identified  processing \egret data
of the CGRO viewing period (VP) 423 (June 20-30, 1995).
The source is located near the Galactic plane
in a field centered at Galactic coordinates $l = 27.31^{\circ}$ and
$b = +1.04^{\circ}$ with an
elongated 
 error box (99\% confidence)
of major axis $\sim 1.4^{\circ}$ and minor axis $\sim 0.8^{\circ}$.
The average \ggg-ray flux above 100~MeV for the whole VP~423 is
$\Phi = (3.3 \pm 0.7)\cdot 10^{-6} \rm \, ph \, cm^{-2} \, s^{-1}$.
The strongly time variable nature of the \ggg-ray emission  from \egr is
evident from the \egret light curve shown in Fig.~1.
As can be seen in Fig.~1, EGRET pointed in the direction of \egr
about 16 times during a period of 4 years. Each exposure was typically
1-2 week long. 
The source was not detected  by EGRET 
during all of the previous observations  except during
VP~334.0 and the combined VPs 421 and 422 with detections above the
$3\,\sigma$ level.
A $\chi^2$ test for variability  yields a probability of
0.0011
that all the EGRET observations of \egr
are consistent with a constant flux.
What makes \egr remarkable is the intensity level reached during the
\ggg-ray flare, with a  peak flux above 100~MeV of
$ (4.0 \pm 1.1)\cdot 10^{-6} \rm \, ph \, cm^{-2} \, s^{-1}$
reached during the last 3.5-day interval of VP~423.
The \ggg-ray luminosity for isotropic emission corresponding to the
peak flux is $L_{\gamma} \simeq 7.2 \times 10^{34} \, d_{kpc}^2 \,
\rm erg \, s^{-1}$, where $d_{kpc}$ is the source distance in kpc.
Fig~1(b) shows the details of the VP~423 EGRET observations
for three 3.5-day intervals.
 This is among the most intense \ggg-ray transients
ever detected, with a \ggg-ray  flux comparable with  that  of
 the Geminga pulsar
and  of  AGN flare   peak intensities as for
3C~279 ($z=0.538$) 
(Kniffen \etal\ 1993)
 and 0528+134 ($z= 2.07$)
(Mukherjee \etal\ 1996).
The \ggg-ray flare in June 1995 is evident in Fig.~1(a), and the
\ggg-ray flux substantially decreased 
before the
next EGRET pointing at the source in late  September 1995.
We notice that prior to June 1995 the source was detected 
only once with marginal significance ($\sim 3.2\sigma$)
during the period July 18-25 1994 (VP~334).
Fig.~1(b) clearly shows the rapid time variability within a 3.5-day timescale
interval. EGRET detected during VP~423 the rising part of a flare
that presumably reached a \ggg-ray flux level even higher than the 
peak flux of $\sim 4\cdot 10^{-6} \rm \, ph \, cm^{-2} \, s^{-1}$.
The EGRET spectrum above 30~MeV during the peak emission
 is consistent with a  power-law  photon spectrum of the form
$ F(E)=k(E/E_o)^{-\alpha}$
with $k=(3.2 \pm 0.6) \times  10^{-9} \rm \,  ph \, cm^{-2} \, s^{-1}
\,  MeV^{-1}$,
$\alpha=2.09 \pm 0.18$ and $E_o=288$~MeV.

Analysis of COMPTEL data for VP 423 in the standard energy intervals 0.75-1
MeV, 1-3 MeV, 3-10 MeV and 10-30 MeV did not show an excess in the maximum 
likelihood skymaps at the position of \egrp.  The $2\,  \sigma$
 upper limit
for the  energy interval 10-30~MeV is 
$3.5 \times 10^{-5} \rm \, ph \, cm^{-2} \, s^{-1}$, 
and  for the most sensitive interval of 1-3~MeV,
$ 3.2 \times  10^{-4} \rm \, ph \, cm^{-2} \, s^{-1}$.
These upper limits are roughly consistent 
with the extrapolation of the EGRET spectral fit for energies above 100 MeV,
as shown in Fig.~2.
We also searched the COMPTEL
maximum likelihood maps of CGRO phase 1-4 periods
in the four standard energy intervals for excess 
emission at the position of \egrp. It is interesting to note that single 
source-like features 
appear  during  VP 334.0 (1-3 MeV, 3-10 MeV) and
during 
VP 302.3 (3-10 MeV) with a significance between 3.0 and 3.7 $\sigma$. However, 
detections cannot be claimed, since in the standard COMPTEL maximum likelihood 
maps the underlying Galactic diffuse emission has not yet been included as a 
background component. A sufficiently accurate model of the diffuse emission
in the COMPTEL energy range
is not yet available. A proper modelling  of the background
component most likely  will reduce the detection significance for
VP~334.0 and VP~302.3.

BATSE can detect hard X-ray  emission from transient sources
with a daily sensitivity of $\sim 75$~mCrab (20-100~keV)
(Harmon \etal\ 1993).
We determined that no  hard X-ray emission from
the direction of \egr was detected  during the 
VP~423 interval. For the whole duration of VP~423, the
$3\sigma$ upper limit in the 20-100~keV energy band is 
$\sim 24$~mCrab.
Fig.~2 shows the multi-instrument  CGRO spectrum
of \egr during VP~423.

\vskip .2in
\section{Search for counterparts}

Fig.~3  shows the radio sources
in the field of \egr
as determined by the Galactic plane survey at 20~cm (Helfand \etal\ 1992).
The EGRET error box of \egr is marked in Fig.~3 by the solid elliptical curve.
We note the presence of several double-lobe sources
(presumably extragalactic jet sources with 
jet axis pointing away from the line of sight),
the radio pulsars PSR~B1831--04 and PSR~B1834--04 
(Taylor, Manchester \& Lyne 1993), and the
center-filled supernova remnant SNR27.8+0.6 (Reich \etal\ 1984).
No spectrally-flat bright radio
source with blazar characteristics  is within the 99\% confidence
error box of \egrp.
We observed the brightest radio source in the error box
(27.920+0.977, source `A' of Fig.~3) at the Green Bank radio interferometer
at 2.25 and 8.3 GHz during the period December 14 1995 - 
January 13 1996. 
We determined that this source has a constant flux
($\sim 500~$mJy at 2.2~GHz)
and  a  steep radio spectral index  $\alpha_r \simeq -1$.
The inferred radio flux at 5~GHz is $S_5 \simeq 0.25$~Jy.
Other radio sources within the error box  are even fainter than
27.920+0.977.
These radio properties are quite different
from those of blazars associated with high confidence with 
\egret sources
(Montigny \etal\ 1995, Thompson \etal\ 1995, M96).
Typically bright \ggg-ray blazars are strong  
($S_{5} \goe 1$~Jy) and  spectrally flat ($\alpha_r \goe -0.5$)
radio sources.

 PSR~B1831--04 and PSR~B1834--04
appear to be radio pulsars with  no anomalous emission.
Their estimated distances are 2.3 and 4.6~kpc, respectively
(Taylor \etal\ 1993).
Their relatively small spindown luminosities 
($10^{32.5}$ and $10^{33.1} \, \rm erg \, s^{-1}$)
are well below the  \ggg-ray luminosity of \egr
put at their distances (even for 100\% efficiency of conversion of
spindown power into gamma-ray emission and a  10\% \ggg-ray beaming factor).

SNR27.8+0.6 is estimated to be at a $\sim 2$~kpc
distance and of age 35,000-55,000 years
(Reich \etal\ 1984). No pulsar 
 is known to be associated with the remnant which appears to be of a
center-filled plerionic type from its radio extended and core
emission.

We determined that three 
X-ray sources in the ROSAT all-sky survey (RASS) database
are within the 99\% confidence error box of \egrp. 
We list here  the names specifying their coordinates (equinox J2000):
RXSJ183745.2-044829, RXSJ183824.1-045034, RXSJ183747.4-035951.
None of these 
is associated with the radio sources of 
Fig.~3.
For a typical countrate of $0.025 \rm \, cts \, s^{-1}$
 and  column density $N_H = 1.23 \cdot 10^{22}
\rm \,  cm^{-2}$, the X-ray energy flux is estimated to be 
$5.3\cdot 10^{-12} \, \rm erg \, cm^{-2} \, s^{-1}$  (0.1-2.4~keV band)
for an assumed photon index of --2 (Voges 1996).
The source
RXSJ183745.2--044829 is coincident (5 arcsecs) with the bright F2 star
($m \simeq 7$) HD~171954.
The other two RASS sources  have no known counterparts.
We also checked that no IRAS sources are 
within the 99\% confidence
 error box of \egrp.

\vskip .2in

\section{Discussion}

We consider first the possibility that \egr is an extragalactic source
of the type detected by EGRET at high Galactic latitudes.
Our motivation is that
the \ggg-ray time variability and flux levels of \egr are  similar to
those occasionally detected from AGNs.
There are four blazars which EGRET has observed with \ggg-ray fluxes as large
as \egr (3C 279, PKS 1622-297, CTA 26, \& PKS 0528+134).
The solid angle subtended by
the  Galactic plane sky area within $\pm 2^\circ$ of Galactic latitude
is $\sim 0.4$~steradians.
The probability  of finding a {\it bright} \ggg-ray blazar 
 is in this case $  \sim 4\times 0.4/4\pi \simeq 0.1$.
However, we note that all bright blazars with a peak
\ggg-ray flux larger than 
$10^{-6}$~ph~cm$^{-2}$~s$^{-1}$ (E$>$100 MeV) have an average 5 GHz radio flux in
excess of 1 Jy (M96).
From the lack of such a counterpart for \egrp, the probability
that this source  is a blazar is substantially smaller than this estimate.

Indeed, if the same fraction $f_q$ of spectrally-steep AGNs with 
$S_5 \loe 0.25$~Jy were producing \ggg-ray flares as for radio-loud 
and spectrally-flat AGNs with $S_5 \goe 1$~Jy ($f_l \sim 0.1$),
EGRET would have revealed a \ggg-ray sky substantially different from
the observed one.
This  can be quantified because 
of very strong evidence for a correlation
between the peak \ggg-ray flux of blazars 
  and the average 5~GHz radio flux (M96). 
 EGRET has detected ten blazars with a peak 
\ggg-ray flux above $10^{-6} \, \rm ph \, cm^{-2} \, s^{-1}$ 
at E$>$100 MeV, and all of these sources  have an average 
$S_5 >  1$~Jy.
We deduce that
\ggg-ray blazars with $S_5 \sim 0.25$~Jy (as source `A')
are at least a factor of 4 (at 90\% confidence) less common than
EGRET blazars with $S_5 \sim 1$~Jy. Also, the great majority of
 blazars detected by EGRET and simultaneously monitored in the radio
are spectrally-flat.
Only two blazars  out of a total of  41 sources
 have indices $\alpha_r \sim  -0.7$ from 
non-simultaneous radio measurements (M96).
Therefore, \ggg-ray blazars with steep indices (as source `A')
are at least a factor of 8 (at 90\% confidence) less common than
EGRET blazars with flat indices.
The probability that
GRO J1838--04 can be identified with source `A' as a blazar 
is less than 0.003. Identification of a blazar with the weaker
radio sources  of 
Fig.~3 is even less likely.
EGRET observations require $f_q \loe 0.03 \, f_l$.

 We note that at the \ggg-ray
 flux level near $10^{-6} \, \rm ph \, cm^{-2} \, s^{-1}$,
there is no lack of sensitivity  for EGRET observations near
the Galactic plane.
 Thus, we conclude 
(99.7\% confidence) that GRO J1838--04 is
not a  bright \ggg-ray  blazar 
 of the kind usually detected by EGRET.
We cannot exclude the possibility that a new type of \ggg-ray AGN of
unusual properties is the counterpart of \egrp.
However, based on five and one half years of EGRET observations
of AGNs, it is highly improbable that a first detection
of such a new \ggg-ray AGN would occur near the  Galactic center.
 Most likely, we are 
dealing with a Galactic phenomenon.

  In principle, 
both binary and
isolated compact objects in the Galaxy  might 
produce variable \ggg-ray emission. Mechanisms for binary systems
include:
(1)  special accretion  and radiation processes  onto the
surface of a  neutron star or  black hole, 
(2) variable  shock-powered synchrotron
 emission from a relativistic pulsar wind termination shock in a
binary system or supernova remnant (e.g., 
Tavani \& Arons, 1997);
(3) inverse Compton `glowing' of  relativistic particle winds (e.g.,
a pulsar wind) subject to cooling in time variable  photon
background baths (as close to periastron of a  massive binary) 
(Tavani 1995);
(4) particle acceleration and radiation in non-relativistic colliding winds of
 a massive binary.
Isolated compact objects might produce gamma-rays because of:
(5)
(6) internal or quake-driven outburst activity from a neutron star.
Model (1)  can be relevant for a transient  X-ray  source
 showing sporadic accretion episodes. 
Among the brightest X-ray transients
detected by BATSE above 100 mCrab, two sources  show
relativistic  jet radio emission,
 GRS~1915+105 (Mirabel \& Rodriguez 1994), and GRO~J1655--40
(Hjellming \& Rupen 1995; see also
Harmon \etal\ 1995, Foster \etal\ 1996).
A gamma-ray flare can be produced by  transient acceleration
processes along the jet in a hypothetical source similar to GRS~1915+105 and
GRO~J1655-40.
However,  the lack of significant hard X-ray  emission from
the direction of \egr as determined by BATSE argues against the
existence of an object similar to known X-ray transients with jet emission.
Both GRS~1915+105 and GRO~J1655--40 
 show hard X-ray outbursts of intensity one-two
orders of magnitude larger than the BATSE upper limit for
\egrp. 
Models (2-4)  based on optically thin emission
are in principle plausible, since
 they refer to a population
of binaries which might exist in the Galactic plane
and to physical processes known to occur in pulsar binaries
(Tavani 1995).
We note that  changes in the mass outflow or orbital parameters
influencing the high-energy emission
of  Galactic binaries can take place
within a few weeks/months. However, in the absence of a  binary
counterpart of \egrp, this interpretation is at the moment hypothetical.
Model (5) requires the existence of a \ggg-ray pulsar with extreme
time-variable magnetospheric emission.
The low-power radiopulsars known in the \egr field, 
PSR~B1831--04 and PSR~B1834--04 
are  not known for any anomalous emission or glitching activity.
Their spindown luminosities are smaller by one-two orders of magnitude
than the required \ggg-ray luminosity. Any new transient
\ggg-ray pulsar at the distance
$d_{kpc}$ is required from our observations
 to have a spindown luminosity larger than
$\sim 7\cdot 10^{35} \, d_{kpc}^2 \, b \, (\varepsilon/0.1)^{-1} \,
\rm erg \, s^{-1}$, where $\varepsilon$ is the efficiency of 
spindown conversion into \ggg-rays and $b$ the \ggg-ray beaming factor.
Model (6) requires outburst conditions and durations  never observed before in
compact stars. 
We notice that at the distance of 2~kpc, a compact object originating
from  SNR27.8+0.6  and moving with a velocity of $\sim 330
\rm \, km \, s^{-1}$ for 50,000 years
 would reach an angular distance of $\sim 0.5^{\circ}$
from the remnant's  center  corresponding to 
 the centroid of the \ggg-ray error box of \egrp.

\vskip .2in
\section{Conclusions} 

We determined that \egr is a strongly variable \ggg-ray
source
and that no blazar counterpart exists
within its 99\% confidence error box.
If \egr is of extragalactic origin,
then
 it shows a new manifestation of
active galaxies. Any 
counterpart is shown here to have radio properties
substantially different from those of blazars detected by EGRET.
The lack of EGRET detections  of low radio luminosity 
(and non-spectrally-flat) AGNs
strongly constrains any extragalactic origin of \egrp.
A Galactic origin of \egr is not as yet supported 
by 
 plausible counterparts.
Additional radio, X-ray and   optical observations  
of the \egr error box are required to gain information
on  its  counterpart.  
A search for binary systems and energetic pulsars in the \egr error
box will be valuable in constraining the nature of the source.

Our results suggest the existence of
a new  class of \ggg-ray sources in addition to
isolated pulsars and radio-loud blazars.
A search for transient \ggg-ray  sources  near the Galactic
plane by current and future high-energy missions is strongly
encouraged.

\vskip .2in
{\small
We thank D. Helfand for
exchange of information
on unpublished data of  Galactic plane radio surveys
and M. Ruderman for discussions.
We are grateful to W.~Voges for his analysis of RASS data. 
 Research  partially supported by the NASA CGRO Guest
Investigator Program (grant NAG~5-2729).
Radio astronomy at the Naval Research Laboratory is supported by the Office
of Naval Research.}
 
\newpage
\vskip .5in
\section{References}

\rref{Foster, R.S., Waltman, E.B., Tavani, M.,
Harmon, B.A. \& Zhang, S.N., 1996, ApJ, 467, L81}

 \rref{Harmon, B.A., \etal, 1993  in {\it 2nd  Compton Observatory
 Symposium}, AIP Conf. no. 304, p. 210}

\rref{Harmon, B.A., \etal,  1995, Nature, 374, 703}

 \rref{Hartman, R.C., \etal, 1996,  ApJ, 461, 698}

\rref{Helfand, D.J., \etal, 1992, ApJS, 80, 211}


\rref{Hjellming, R.M. \& Rupen, M.P., 1995, Nature, 375, 464}

\rref{Hunter, S., \etal, 1996, submitted to ApJ}


\rref{Kniffen, D., \etal, 1993, ApJ, 411, 133}



\rref{Mattox, J.R., Schachter, J., Molnar, L., Hartman, R.C.,
Patnaik, A.R., 1996, submitted to ApJ (M96)}


\rref{Mirabel, I.F., \& Rodriguez, L.F., 1994, Nature 371,  46}

\rref{Montigny, C.V., \etal, 1995, ApJ, 440, 525}

 \rref{Mukherjee, R. \etal, 1996, ApJ, 470, 831}
 

 
\rref{Ramanamurthy, P.V., \etal, 1995, ApJ, 450, 791}

\rref{Reich, W., Furst, E. \& Sofue, Y., 1984, A\&A, 133, L4}

\rref{Ruderman, M., 1996, private communication}
 
 \rref{Tavani, M., 1995, in
{\it The Gamma-Ray Sky of SIGMA and GRO},
eds. M. Signore, P. Salati, G. Vedrenne (Dordrecht: Kluwer), p. 181}
 
 \rref{Tavani, M. \& Arons, J., 1997,  ApJ, in press}
 
 

 \rref{Tavani, M., \etal, 1997,  to be submitted to ApJ}


\rref{Taylor, J.H., Manchester, R.N. \& Lyne, A., 1993, ApJS,  88, 529}

\rref{Thompson, D.J., \etal, 1995, ApJS, 101, 259}

\rref{Thompson, D.J., \etal, 1996, ApJS, 107 (Nov. 1996)}

\rref{Voges, W., 1996, private communication}
 

\newpage

\centerline{Figure Captions}

\nn
{\bf Figure 1 (a)} --
Time history of \ggg-ray flux detected by \egret from the field
containing \egrp. TJD is the truncated Julian date (JD), TJD=JD--2,400,000.0. 
$1 \sigma$ flux  errors  and $2 \sigma$
upper limits are reported, the upper limits being shown as downward
arrows.
The time interval is from April 1991 through the end of September 1995.
EGRET data for VPs 421 and 422 have been combined. 
{\bf (b)} Time history for the VPs 421, 422, 423 and 429,
spanning a time interval from 
June 6 until September 27, 1995.
EGRET data for VP~423 have been divided into three 3.5-day intervals
($1 \sigma$ flux  error bars are shown).
 

\vskip .1in
\nn
{\bf Figure 2} -- 
Multi-instrument CGRO  spectrum of \egr
during the period of maximum \ggg-ray emission 
(VP~423, June 20-30, 1995); $2\sigma$~upper limits are marked
as downward arrows.

\vskip .1in 
\noindent
{\bf Figure 3} --
Radio sources of the Galactic plane  survey at 20~cm
(Helfand \etal\ 1992) with the 99\% confidence EGRET error box
of \egr marked by the solid ellipse.
The diamond's size  is proportional to the radio flux at 20 cm:
source A, 738~mJ;
source B, 264+99~mJ;
source C, 23+27~mJy;
source D, 83+377~mJy;
source E, 69~mJy;
source F, 68+41~mJy.
Crosses indicate the positions of the  radiopulsars
PSR~B1831-04 and PSR~B1834-04.
The dashed line circle approximates the extension of the  
supernova remnant G27.8+0.6 at 2.6 GHz (Reich \etal\ 1984).
Filled circles give the positions of the RASS sources.
No spectrally-flat radio-loud AGN is detected 
 within the error box of \egrp.
 

\newpage
\epsfxsize=6.5in
\epsfysize=4.in
\epsffile{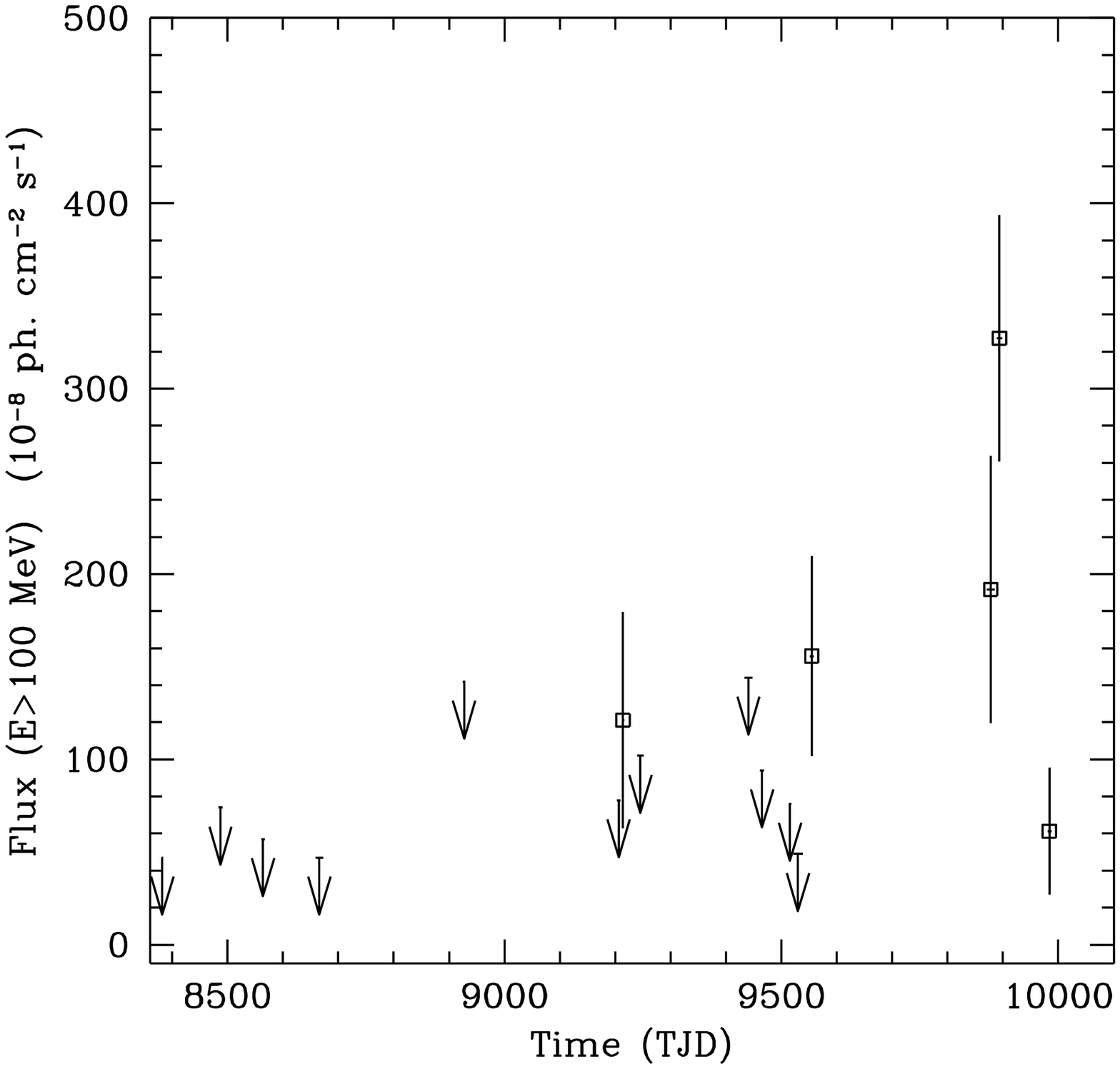}
\cen{Fig.~1 (a)}

\epsfxsize=6.5in
\epsfysize=3.5in
\epsffile{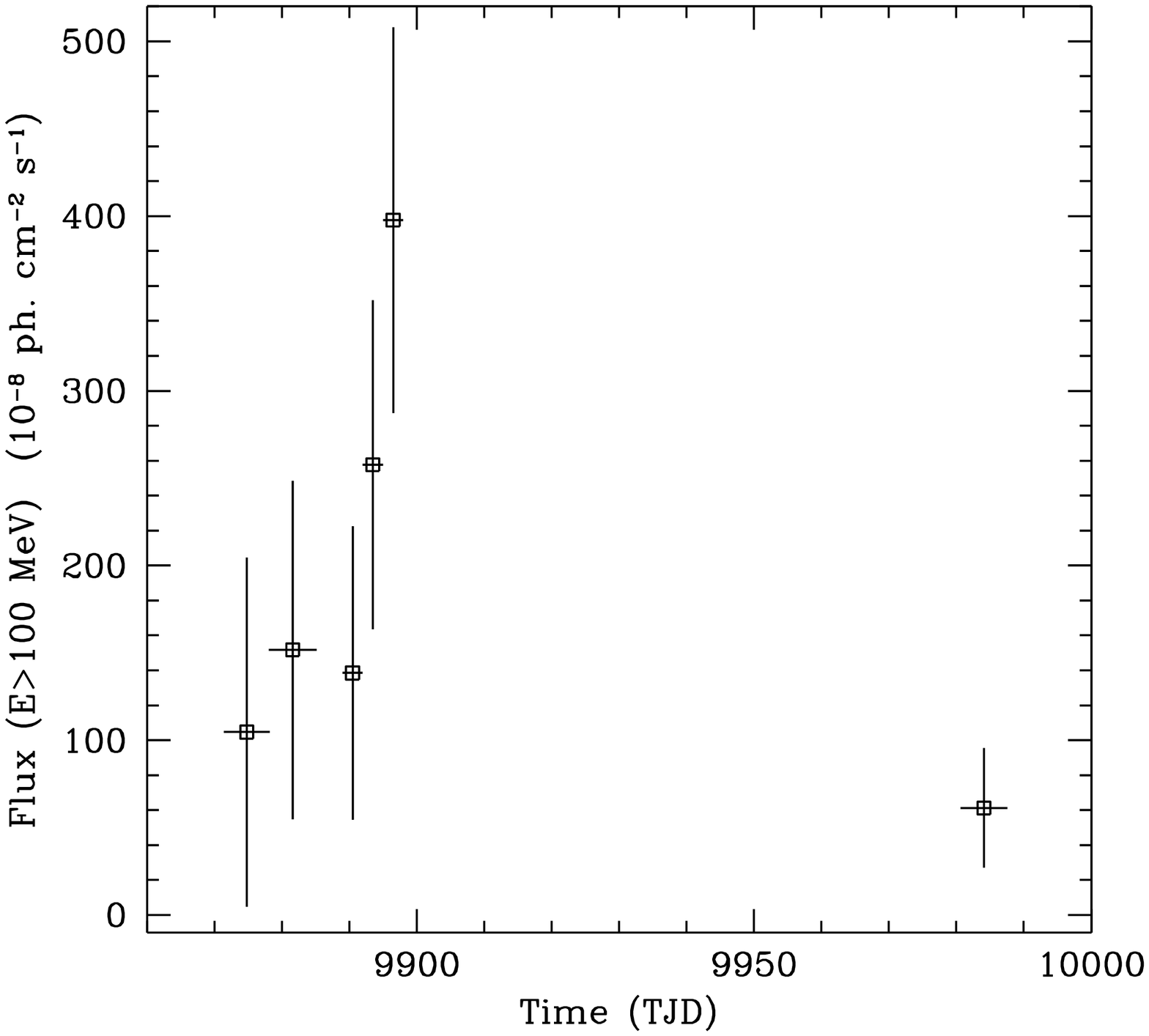}
\cen{Fig.~1 (b)}

\newpage

\epsfxsize=6.5in
\epsfysize=8.in
\epsffile{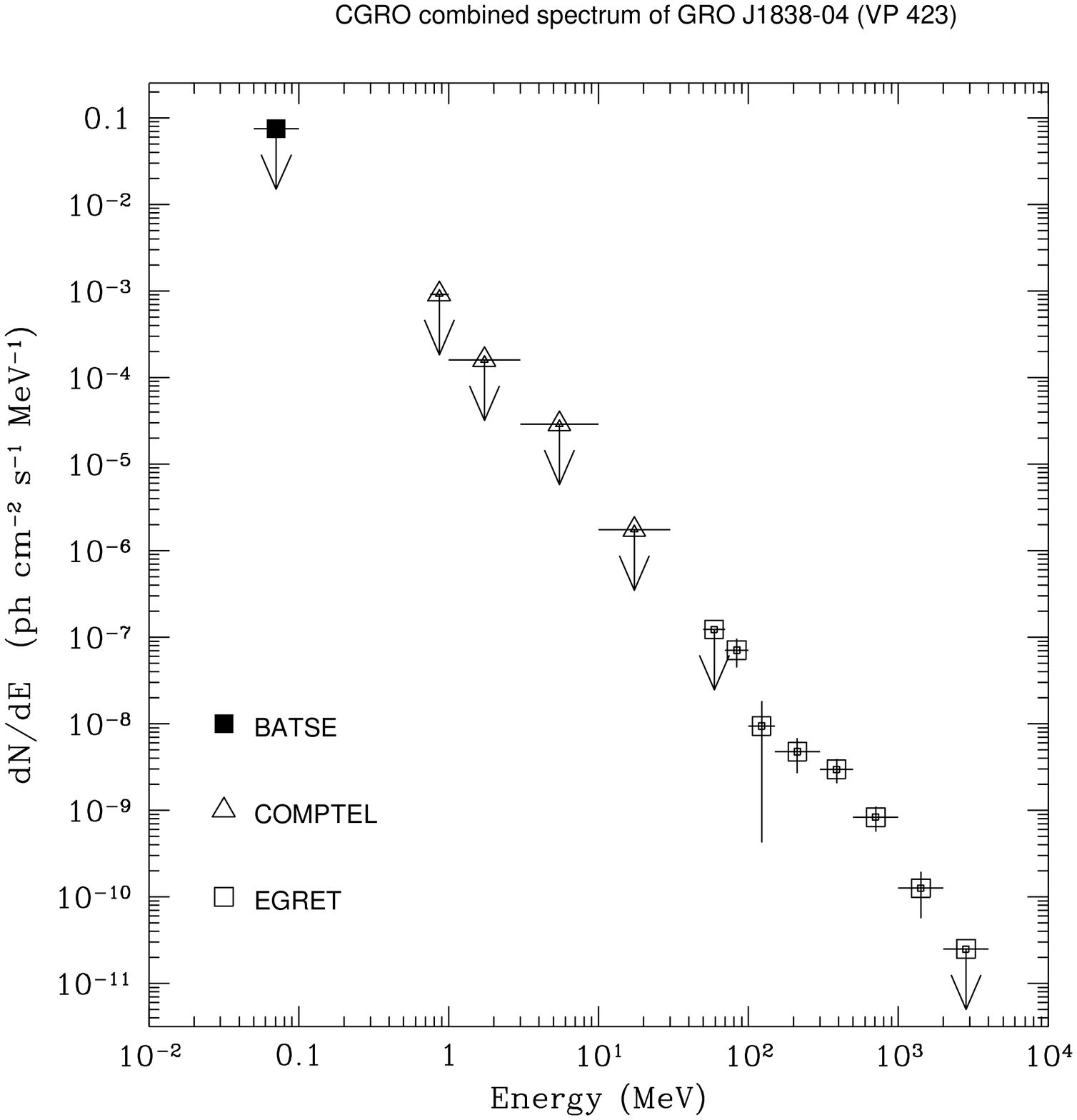}
\cen{Fig.~2}

\newpage

\vspace*{3.in}
\epsfxsize=8.5in
\epsfysize=7.in
\epsffile{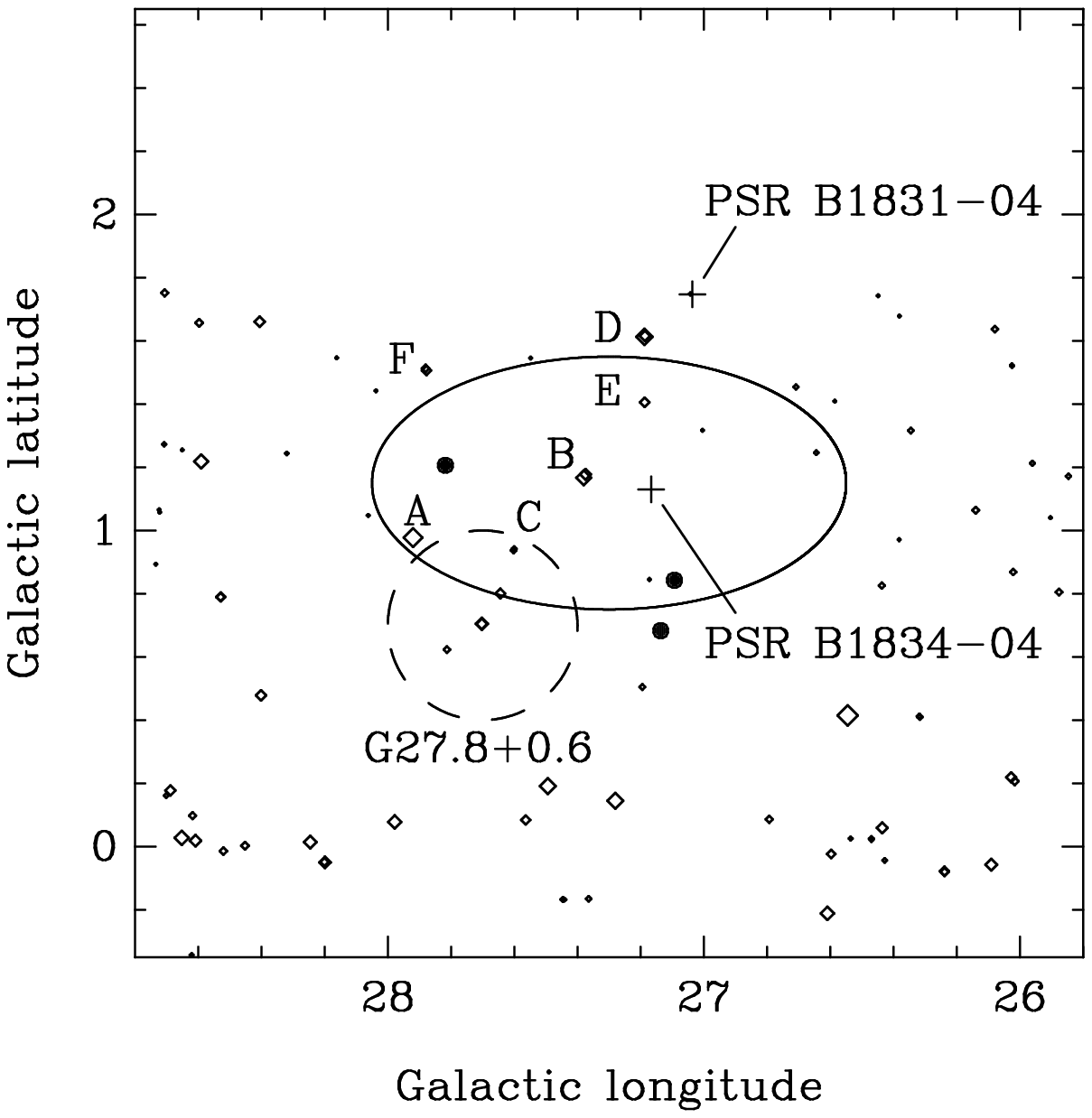}
\vspace*{-3.in}
\cen{Fig.~3}

\end{document}